\documentclass[draft]{agujournal20192}
\usepackage{url} 
\usepackage{lineno}
\usepackage[inline]{trackchanges} 
\usepackage{soul}
\draftfalse

\usepackage{graphicx}
\usepackage{amsmath}
\usepackage{amssymb}
\usepackage{times}
\usepackage[letterpaper,margin=1in]{geometry}
\usepackage{caption}
\usepackage{array} 
 
\newcommand{\V}[1]{\mathbf{#1}} 
 
\newcommand\Alfven{Alfv\'en }
\newcommand\Alfvenic{Alfv\'enic }
\newcommand{\tabref}[1]{Table~\ref{#1}}
\newcommand{\figref}[1]{Figure~\ref{#1}}

\usepackage{color}

\newcommand{\blue}[1]{\textcolor{black}{#1}}
\newcommand{\newblue}[1]{\textcolor{black}{#1}}

\journalname{JGR: Space Physics}

\begin{document}

\title{The Importance of Electron Landau Damping for the Dissipation of Turbulent Energy in Terrestrial Magnetosheath Plasma}

\authors{A. S. Afshari\affil{1},  G. G. Howes\affil{1}, C. A. Kletzing\affil{1}, D. P. Hartley\affil{1}, S. A. Boardsen\affil{2,3}}

\affiliation{1}{Department of Physics and Astronomy, The University of Iowa, Iowa City, IA USA}
\affiliation{2}{Goddard Planetary Heliophysics Institute, University of Maryland, Baltimore, MD, USA}
\affiliation{3}{NASA/GSFC, Greenbelt, MD, USA}

\correspondingauthor{Arya S. Afshari}{arya-afshari@uiowa.edu}

\begin{keypoints}
\item Electron Landau damping plays a ubiquitous role in dissipating \blue{\Alfvenic}turbulent energy in the magnetosheath.
\item Electron Landau damping signatures correspond to the wave energy flux direction, implying the dissipation of locally imbalanced turbulence.
\item Electron Landau damping sometimes plays a dominant role in dissipating turbulent energy in the magnetosheath. 
\end{keypoints}

\begin{abstract}
    Heliospheric plasma turbulence plays a key role in transferring the energy of large-scale magnetic field and plasma flow fluctuations to smaller scales where the energy can be dissipated, ultimately leading to plasma heating.  High-quality measurements of electromagnetic fields and electron velocity distributions by the Magnetospheric Multiscale (MMS) mission in Earth's magnetosheath present a unique opportunity to characterize plasma turbulence and to determine the mechanisms responsible for its dissipation. We apply the field-particle correlation technique to a set of twenty MMS magnetosheath intervals to identify the dissipation mechanism and quantify the dissipation rate. It is found that 95\% of the intervals have velocity-space signatures of electron Landau damping that are quantitatively consistent with linear kinetic theory for the collisionless damping of kinetic Alfv{\'e}n waves. About 75\% of the intervals contain asymmetric signatures, indicating a \emph{local imbalance} of kinetic Alfv{\'e}n wave energy flux in one direction along the magnetic field than the other. \blue{About one third} of the intervals have an electron energization rate with the same order-of-magnitude as the estimated turbulent cascade rate, suggesting that electron Landau damping plays a significant, and sometimes dominant, role in the dissipation of the turbulent energy in these magnetosheath intervals.\\
\end{abstract}

\section{Introduction}
Turbulence is a fundamental, yet poorly understood, process that transfers the energy of chaotic plasma flows and electromagnetic fields into the energy of the plasma particles, either as heat or some other form of particle energization, representing a grand challenge problem in heliophysics. Within the heliosphere, turbulent plasma heating and particle acceleration play a key role in governing the flow of energy, impacting the mesoscale and macroscale evolution of the heliospheric environments comprising the coupled solar-terrestrial system and connecting the solar corona, solar wind, and Earth’s magnetosphere \cite{NRCspace:2013}. A long-term goal in space physics and astrophysics is to understand and predict how turbulence couples the large-scale plasma conditions and evolution to the microphysical heating and particle energization. Before this can be accomplished, however, it is essential to identify and quantitatively characterize the microphysical kinetic processes that govern the energization of particles through the dissipation of turbulence, and here we report the achievement of a critical milestone in that effort. 

For most space and astrophysical plasmas, the small-scale end of the turbulent cascade---where turbulent energy is ultimately converted to plasma heat or non-thermal particle energization---is dominated by weakly collisional plasma kinetics \cite{Schekochihin:2009, Howes:2017b}. The proposed particle energization mechanisms under these weakly collisional conditions fall into three broad categories: (i) resonant wave-particle interactions, such as Landau damping \cite{Landau:1946, Dobrowolny:1985, Leamon:1999, Howes:2008, Schekochihin:2009}, transit-time damping \cite{Barnes:1966}, or cyclotron damping \cite{Isenberg:1983, Hollweg:2002}; (ii) nonresonant wave-particle interactions, such as stochastic ion heating \cite{Johnson:2001,Chandran:2010}, or magnetic pumping \cite{Lichko:2017,Lichko:2020}; and (iii) dissipation in coherent structures, such as collisionless magnetic reconnection occurring in small-scale current sheets \cite{Ambrosiano:1988,Osman:2011,Zhdankin:2015}. To develop a predictive capability for particle energization in plasma turbulence, we must first identify the dominant particle energization mechanisms and characterize how they vary with the plasma and turbulence parameters found in different space and astrophysical environments. 

Plasma turbulence has been studied using \emph{in situ} measurements in the solar wind \cite{Tu:1995, Bruno:2013, Kiyani:2015} and planetary magnetospheres \cite{Saur:2004, vonPapen:2014, Hadid:2015, Tao:2015, Ruhunusiri:2017}. These studies have shown that the turbulent power spectrum is embedded in the solar wind and changes through interactions with magnetized bodies. Earth's magnetosheath---the region downstream of the bowshock, bounded by the bowshock and the magnetopause---has been shown to exhibit a power law scaling of the magnetic field power spectral density (PSD) which is indicative of turbulence \cite{Chasapis:2015, Hadid:2015, Voros:2016, Huang:2017}. The regions of the magnetosheath downstream from the bowshock, near the magnetopause, provide the conditions necessary to study well-developed turbulence and the turbulent dissipation mechanisms that transfer the energy from the turbulent fields to the plasma particles \cite{Huang:2017,Hadid:2018,Chen:2019}.

Although a number of studies have found observational evidence consistent with the predictions of some of the dissipation mechanisms \cite{Kasper:2008, Osman:2011, He:2015a}, to date there have been few studies that directly identify a particular mechanism or quantitatively assess its contribution to the removal of energy from the turbulent cascade. Using spacecraft observations in the Earth's turbulent magnetosheath plasma, a case study by \citeA{Chen:2019} made the first direct identification of electron Landau damping in space plasma turbulence through its unique velocity-space signature using the Field-Particle Correlation (FPC) technique \cite{Klein:2016,Howes:2017a,Klein:2017}. 

Here we exploit the unique, high-cadence measurements of the electromagnetic fields and electron velocity distributions in the Earth's turbulent magnetosheath plasma by the Magnetospheric Multiscale (MMS) mission \cite{Burch:2016} to find that velocity-space signatures of electron Landau damping are nearly ubiquitous in a sample of 20 intervals. We find intervals with symmetric or asymmetric velocity-space signatures of electron Landau damping, and we find that the direction of these signatures, with respect to the background magnetic field, where Landau damping is observed corresponds to the direction of the balanced or imbalanced kinetic Alfv{\'e}n wave (KAW) energy flux. Furthermore, we quantitatively assess the contribution of electron Landau damping to the dissipation of the turbulent cascade, finding that this mechanism dominates the turbulent dissipation rate in about \blue{one third} of the intervals in our study.

This paper is organized in the following manner: in Section 2 we outline the methodology by describing the FPC technique, the physical quantities to be examined, the MMS data sets, and the data analysis procedures; Section 3 contains the results of the FPC technique applied to the MMS data set, the observations of the wave energy flux, and the comparison of the electron energization rates to the theoretically estimated turbulent energy cascade rates; in Section 4 we discuss our results and make a comparison with other turbulent magnetosheath studies; in Section 5 we present our conclusions; Appendices A - D provide further details of our work.


\section{Methodology}

\subsection{The Field Particle Correlation Technique}

The FPC technique in general utilizes the full 3V velocity-space measurements to generate \emph{velocity-space signatures} that are characteristic of the kinetic mechanisms that govern the dissipation of turbulence and consequent energization of particles \cite{Klein:2016, Howes:2017a, Klein:2017, Klein:2020}. Collisions may be neglected on the time scale of the collisionless energy transfer that removes energy from the turbulence, so the Vlasov equation for species $s$ describes the plasma dynamics,
\begin{equation}
\frac{\partial f_s}{\partial t} +{\bf v} \cdot {\bf \nabla} f_s + \frac{q_s}{m_s}\left[ {\bf E} + {\bf v} \times {\bf B}  \right] \cdot  \frac{\partial f_s}{\partial {\bf v}} = 0. 
\end{equation}
The Vlasov equation is multiplied by $m_s v^2/2$ to obtain the time evolution of the 3D-3V phase-space energy density, $w_s(\V{r},\V{v},t)=m_s v^2 f_s(\V{r},\V{v},t)/2$, 
\begin{equation}
\frac{\partial w_s}{\partial t} =  - {\bf v} \cdot {\bf \nabla} w_s - q_s \frac{v^2}{2} {\bf E} \cdot \frac{\partial f_s}{\partial {\bf v}} - q_s\frac{v^2}{2} ({\bf v} \times {\bf B}) \cdot  \frac{\partial f_s}{\partial {\bf v}}.
\label{eq:dwdt}
\end{equation}

When Equation \eqref{eq:dwdt} is integrated over all 3D-3V phase-space, yielding the rate of change of total energy $\mathcal{W}_s$ of species $s$, only the electric field term is non-zero \cite{Howes:2017a}, so any net change in energy is due to the electric field.
Since it is the parallel component of the electric field, $E_\parallel$, that energizes particles in Landau damping, the rate of electron energization by Landau damping can be assessed by taking the unnormalized correlation,
\begin{equation}
C_{E_{\parallel}}({\bf v}, t, \tau) = C\left( -q_e \frac{v_{\parallel}^2}{2}\frac{\partial f_e({\bf r}_0, {\bf v}, t)}{\partial v_{\parallel}}, E_{\parallel}({\bf r}_0, t) \right) = -\frac{1}{N}\sum_{j=1}^{N} q_e \frac{v_{\parallel}^2}{2}\frac{\partial f_e({\bf r}_0, {\bf v}, t_j)}{\partial v_{\parallel}} E_{\parallel}({\bf r}_0, t_j)
\label{eq:cepar}
\end{equation}
where N is the number of electron distribution measurements in the correlation interval $\tau$. The correlation is taken over a sufficiently long correlation interval $\tau$ to average out the oscillatory energy transfer associated with undamped wave motion, exposing the smaller amplitude signal of secular energization \cite{Klein:2016,Howes:2017a}. This secular energy is the net energy transferred between $E_\parallel$ and the particles during the correlation interval. In the application of the FPC technique here, we employ the perturbed electron velocity distribution integrated over the azimuthal angle in cylindrical velocity space $(v_\parallel, v_\perp, \phi)$, given by $\delta f_e(v_\parallel, v_\perp, t)$, to compute the correlation $C_{E_{\parallel}}(v_\parallel,v_\perp, t,\tau)$.  Subsequently,  integrating over $v_\perp$ yields the reduced parallel correlation, $C_{E_{\parallel}}(v_\parallel,t,\tau)$.  

\begin{figure}
\begin{center}
    \resizebox{5.0in}{!}{\includegraphics[width=\textwidth]{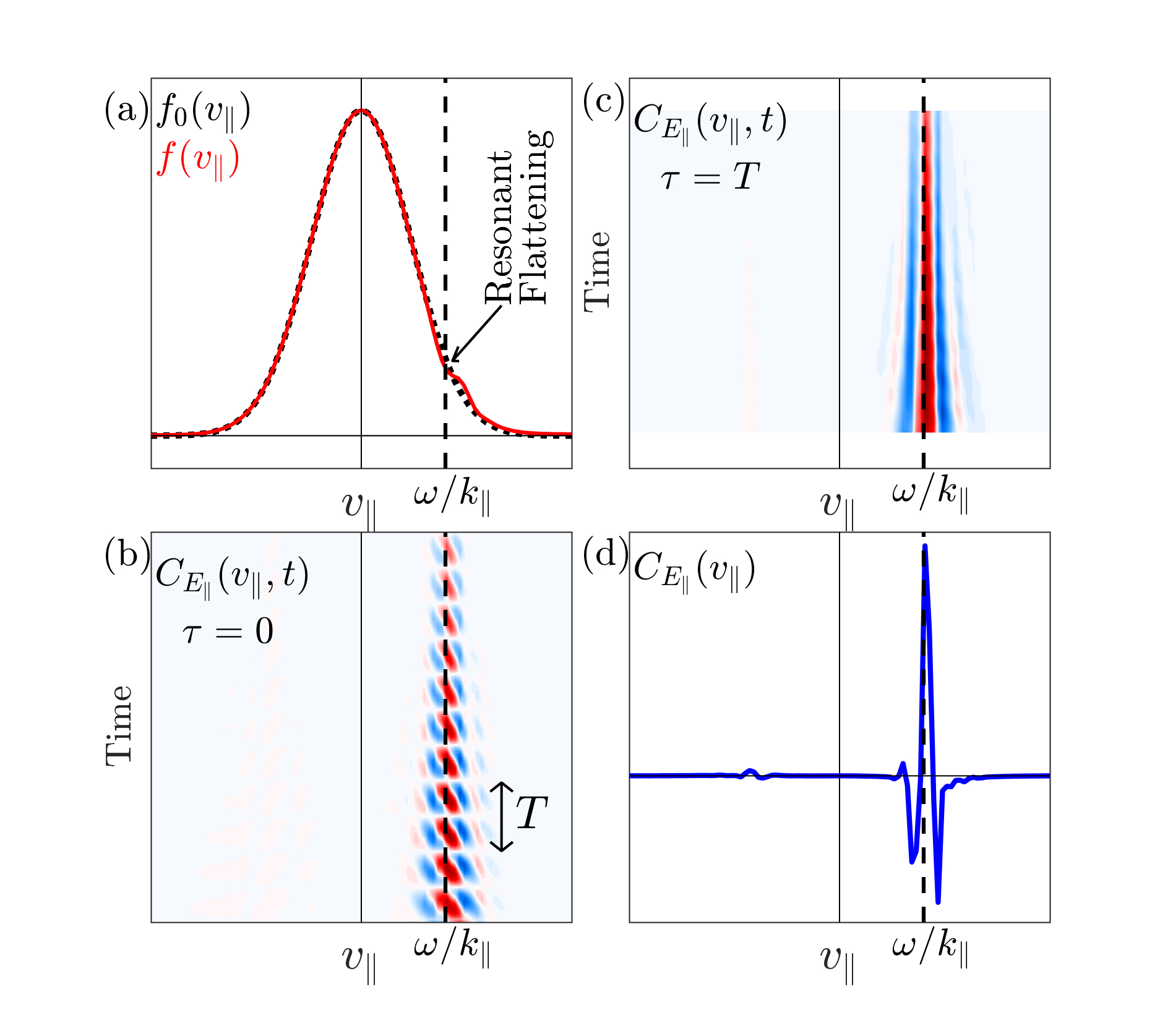}}
\end{center}    
    \caption{\label{fig:Landaudamping_FPC} (a) Landau damping leads to a flattening of the velocity distribution $f(v_\parallel)$ at the resonant phase velocity $v_\parallel = \omega/k_\parallel$. (b) An instantaneous field-particle correlation with $\tau=0$ yields a signature in $v_\parallel$ that oscillates in time. (c) Averaging over an interval equal to the wave period $\tau=T$ leads to a persistent velocity-space signature. (d) The time-averaged correlation $C_{E_\parallel}(v_\parallel)$ exhibits a bipolar signature, crossing from negative to positive at the resonant velocity, a distinguishing characteristic of Landau damping.}
\end{figure}

A timestack plot of the reduced parallel correlation correlation $C_{E_{\parallel}}(v_\parallel,t,\tau)$ as a function of parallel velocity $v_\parallel$ and time $t$, shown in \figref{fig:Landaudamping_FPC}, illustrates the typical velocity-space signature of Landau damping. \blue{This figure is an illustrative diagram of electron Landau resonance and the resulting plots when the FPC technique is applied to it.} Here (a) the perturbation of $f_0(v_\parallel)$, associated with a KAW travelling parallel to the magnetic field with phase velocity $\omega/k_\parallel$ (vertical dashed line), leads to (b) an instantaneous correlation $C_{E_{\parallel}}(v_\parallel,t,0)$ dominated by oscillatory energy transfer in time about the resonant phase velocity $\omega/k_\parallel$ (vertical dashed line).  By taking the average over a sufficiently long correlation interval, here equal to the wave period $\tau=T$, (c) the correlation $C_{E_{\parallel}}(v_\parallel,t,\tau)$ becomes steady in time, and (d) its time average yields the characteristic bipolar form of the velocity-space signature for Landau damping where $C_{E_{\parallel}}(v_\parallel)$ crosses from negative to positive at the resonant velocity  \cite{Klein:2016, Howes:2017a, Howes:2017b, Klein:2017, Howes:2018a, Klein:2020}.  Physically, the net effect of the resonant interaction with $E_\parallel$ is that particles with $v_\parallel<\omega/k_\parallel$ are accelerated to 
$v_\parallel>\omega/k_\parallel$, leading to a net loss of phase-space energy density below the resonance, and an increase above. This leads to the bipolar velocity-space signature seen in \figref{fig:Landaudamping_FPC}(d).

To avoid difficulties with taking velocity-space derivatives in the 2V velocity-space of $f_e(v_\parallel, v_\perp, t)$, the procedure of \citeA{Chen:2019} is followed and the alternative correlation,
\begin{equation}
C'_{E_{\parallel}}(v_\parallel, v_\perp, t, \tau) = \langle q_e v_{\parallel} f_e(v_\|,v_\bot,t) E_{\parallel}(t) \rangle_\tau,
\label{eq:full_cprime}
\end{equation}
in 2V space is calculated. The resulting correlation 
$C'_{E_{\parallel}}(v_\parallel,v_\bot,t,\tau)$ is then \blue{binned in 2V velocity-space with square bin sizes of 10\% the electron thermal speed, thus providing sufficient velocity-space resolution to resolve resonances in the thermal population. Though in the high velocity tails of the electron distribution the counts may be lower as a result of this 10\% $v_{th,e}$ bin size, they are not involved in the dissipation of turbulent energy observed here, thus bearing no outcome on our results.}

This binned $C'_{E_{\parallel}}(v_\parallel,v_\bot,t,\tau)$ is integrated over $v_\perp$ to obtain $C'_{E_{\parallel}}(v_\parallel,t,\tau)$,
\blue{
\begin{equation}
C'_{E_{\parallel}}(v_\parallel, t, \tau) = 2\pi\int dv_\bot v_\bot C'_{E_{\parallel}}(v_\parallel,v_\bot,t,\tau).
\label{eq:reduced_cprime}
\end{equation}
}
After this integration, $C_{E_{\parallel}}(v_\parallel,t,\tau)$ is obtained by computing 
\begin{equation}
C_{E_{\parallel}}(v_\parallel, t, \tau) =  -\frac{v_\parallel}{2} \frac{\partial C'_{E_{\parallel}}(v_\parallel, t, \tau)}{\partial v_\parallel} + \frac{C'_{E_{\parallel}}(v_\parallel, t, \tau)}{2}.
\label{eq:reduced_full_C}
\end{equation}
The time-averaged signature of electron Landau damping (as in \figref{fig:Landaudamping_FPC}(d)) is found by letting $\tau = T$ where $T$ is the full duration of the interval being analyzed.

Integrating $C_{E_{\parallel}}(v_\parallel,\tau)$ over $v_\parallel$ yields the rate of change of electron spatial energy density due to $E_\parallel$ at the point of observation, given by 
\begin{equation}
    \left(\frac{\partial W_e(\V{r}_0,t)}{\partial t}\right)_{E_\parallel}=\int d v_\| C_{E_\parallel}(v_\parallel,t).
\label{eq:eenergization}
\end{equation}
The electron energization rate obtained in Equation \eqref{eq:eenergization} is equivalent to the time average of the contribution to $\V{J}_e\cdot\V{E}$ by $E_\parallel$ (see \ref{appendix:quality_checks} for details and validation).

\subsection{Parallel Poynting Flux}

The parallel Poynting flux, or wave energy flux, indicates the direction of the net flow of electromagnetic energy due to \Alfven waves propagating up and down the mean magnetic field $\V{B}_0$, a quantity not obvious from magnetic and electric field measurements alone \cite{Kelley:1991}.  When the turbulence is \emph{balanced} \cite{Lithwick:2007,Howes:2008}, it means that the \Alfven wave energy flux up (parallel to) the magnetic field is equal, on average, to that down (anti-parallel to) the magnetic field. In this balanced case, the Poynting flux would average to zero.  If electron Landau damping acts to remove energy from the turbulence in this case, we expect the qualitative signature of electron Landau damping to be symmetric about $v_\|=0$, indicating the dissipation of KAWs travelling both parallel and anti-parallel to $\V{B}_0$. However, even in a case of turbulence that is balanced over longer time scales, for relatively short times one may observe \emph{locally imbalanced turbulence}, where the KAW energy flux is greater in one direction along $\V{B}_0$ than the other, leading to a nonzero net Poynting flux up or down the magnetic field.  In this case, for Landau damping of the imbalanced KAWs, we expect an asymmetric signature of Landau damping, with a larger bipolar signature in the same direction in $v_\|$ as the dominant \Alfven wave energy flux. The wave energy flux is calculated as the component of the Poynting vector parallel to \textbf{B}$_0$,
\begin{equation}
S_\|(t)=\frac{1}{\mu_0}\left[\delta \V{E}_\bot(t)\times \delta\V{B}_\bot(t)\right] \cdot \hat{\V{b}}_0,
\label{eq:s_para}
\end{equation}
where $\delta \V{E}_\bot(t)$ and $\delta \V{B}_\bot(t)$ are the electric and magnetic field fluctuations perpendicular to $\V{B}_0$ \cite{Kelley:1991} and $\hat{\V{b}}_0=\V{B}_0/|\V{B}_0|$ is the unit vector in the direction of the mean magnetic field. The local imbalance of the turbulence is assessed by considering the normalized Poynting flux,
\begin{equation}
\hat{S_\|}(t) = \frac{\langle S_\|(t) \rangle_\tau} {(\langle \delta\V{E}_\bot^2(t)\rangle_\tau\langle\delta\V{B}_\bot^2(t)\rangle_\tau)^{1/2}},
\label{eq:s_para_norm}
\end{equation} 
where the angle brackets denote averages over intervals of $\tau$.

\subsection{Turbulent Energy Cascade Model}

The theoretically estimated turbulent energy cascade rate $\epsilon$ in the inertial range from a cascade model \cite{Howes:2008, Howes:2011} is given by
\begin{equation}
\epsilon \sim \frac{\mbox{Energy Density}}{\mbox{Cascade Time}} = \frac{n_0 m_p U_\perp^2}{1/(k_\perp U_\perp)}
= n_0 m_p\left( \frac{2 \pi f}{v_{\bot,sc}}\right) \frac{[\delta \hat{B}_\perp(f)]^3}{(\mu_0 n_0 m_p)^{3/2}}.
\label{eq:eps}
\end{equation}
Here the turbulent energy density includes both kinetic and magnetic contributions, which are assumed equal for \Alfvenic turbulence, and the cascade time is calculated as the eddy turn-around time at MHD scales. The amplitude of the magnetic field fluctuations, $\delta \hat{B}_\perp(f)$, is computed at a frequency $f=0.2$~Hz within the inertial range, computed from the time series as the increment with lag $t=5$ s. \blue{This frequency is chosen as it is in the inertial range for the analyzed intervals.} The Taylor hypothesis \cite{Taylor:1938} is assumed to estimate $k_\perp$ as the factor in parentheses where $v_{\bot,sc}$ is the component (perpendicular to $\V{B}_0$) of the electron plasma flow relative to the spacecraft \cite{Howes:2014}. The Taylor hypothesis has been shown to be generally applicable in the magnetosheath \cite{Chhiber:2018}, and a typical anisotropy $k_\parallel \ll k_\perp$ is assumed \blue{\cite{Sahraoui:2006,Sahraoui:2010b,Roberts:2013,Roberts:2015b}} since the intervals are chosen closer to the magnetopause where the turbulence is expected to be well developed \cite{Huang:2017}. The MHD \Alfven wave eigenfunction is used to estimate $U_\perp =\delta B_\perp v_A/B_0$. 

\subsection{Data Set and Data Analysis}
\subsubsection{Selection of Intervals and Data Set}
MMS magnetosheath intervals are identified through the use of the orbit plots, Fluxgate Magnetometer (FGM) data \cite{Russell:2016}, and Fast Plasma Investigation (FPI) density data and particle energy spectra \cite{Pollock:2016}. The magnetosheath is defined as the layer between the bow shock and the magnetopause, with generally higher temperatures and higher densities than the solar wind and magnetospheric plasmas, and it provides a confined region in which turbulence evolves. Previous work has shown that the plasma turbulence in the magnetosheath is statistically likely to be well developed downstream of the bowshock, closer to the magnetopause, and away from the subsolar point \cite{Huang:2017}.

\begin{figure}
\begin{center}
    \resizebox{5.0in}{!}{\includegraphics[width=\textwidth]{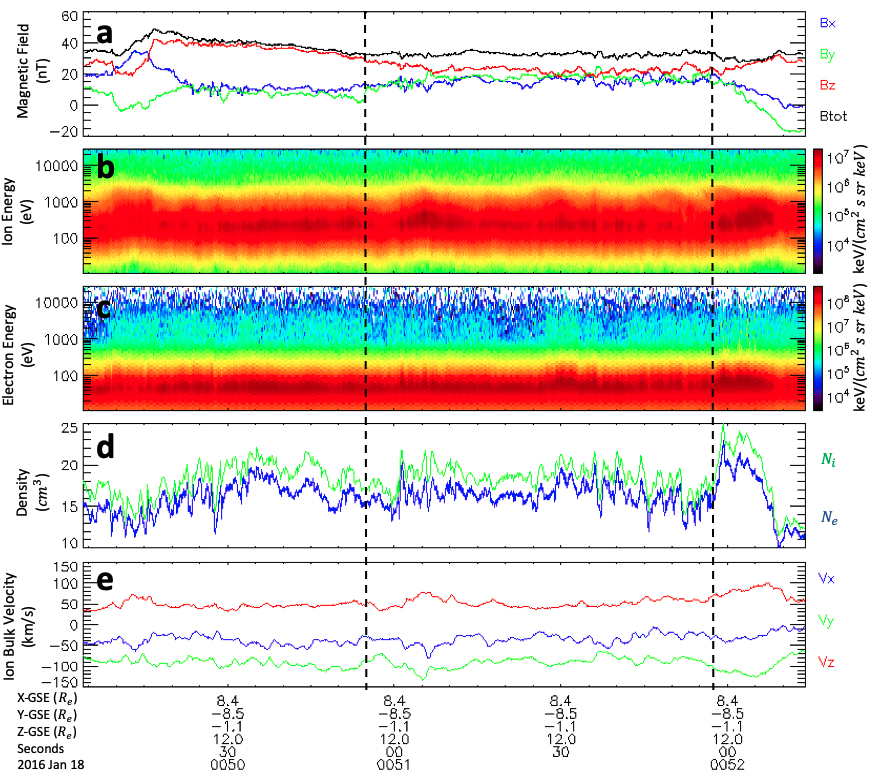}}
\end{center}    
    \caption{\label{fig:interval08_characteristics} An example of one of the turbulent intervals selected for analysis. Interval 08 during 2016-01-18 from MMS 1: (a) the magnetic field, (b) ion energy spectrogram, (c) electron energy spectrogram, (d) electron and ion densities, and (e) ion bulk velocity. The 1 minute and 2 second interval between the dashed vertical lines is chosen for the FPC analysis.}
\end{figure}

During periods in which the MMS spacecraft were in the Earth's magnetosheath, we select twenty intervals with relatively constant magnetic field direction and particle densities, as well as no velocity shears. Time-averaging the magnetic field components over the full duration of the interval $T$ to obtain magnetic Field-Aligned Coordinates (FAC) requires an accurate definition of the ambient magnetic field $\V{B}_0$. This is facilitated when the magnetic field direction is relatively constant with no discontinuities. These intervals with relatively constant magnetic field magnitude and plasma density indicate primarily Alfv{\'e}nic fluctuations \blue{where the fluctuations in the magnetic field $\delta$\textbf{B} are small compared to the magnitude of the ambient magnetic field $|\V{B}_0|$. In these intervals, there is} negligible energy in compressible fluctuations, as typically found in observations of turbulence in the solar wind \cite{Bruno:2013}. A representative example which matches these criteria is given in \figref{fig:interval08_characteristics} when the MMS 1 spacecraft is in the magnetosheath, closer to the magnetopause than the bowshock. The FPC technique requires only single-point measurements of the electromagnetic fields and particle velocity distributions \cite{Klein:2016, Howes:2017a, Klein:2017}, so, for each interval analyzed, we select data from only one of the four MMS observatories. All intervals were checked for the basic characteristics of turbulence, mainly that the trace magnetic power spectrum follows an approximate $-5/3$ frequency scaling at spacecraft-frame frequencies $f_{sc} < 0.5$~Hz, with steepening at higher frequencies. The twenty intervals selected for analysis, including their respective durations $T$ and plasma parameters, are listed in \tabref{tab:sample_details}. 

\subsubsection{Coordinate Systems and Ambient Magnetic Field}
Two coordinate systems are used in this study: Geocentric Solar Ecliptic (GSE) and magnetic Field-Aligned Coordinates (FAC). In the GSE coordinate system, the Earth is at the origin, the positive X direction points from the Earth to the Sun, the positive Z direction points northward out of the ecliptic plane, and the positive Y direction completes the right-handed orthogonal system by pointing from dawn to dusk. In the FAC coordinate system, the magnetic field direction is the relevant axis with the parallel direction ($\|$) being parallel to the ambient magnetic field \textbf{B}$_0$, and the perpendicular direction ($\bot$) being in the plane perpendicular to \textbf{B}$_0$. The ambient magnetic field \textbf{B}$_0$ is defined as the time-averaged components of the magnetic field measurements in GSE from the FGM data. \textbf{B}$_0$ is given by \textbf{B}$_0=\langle B_x(t),B_y(t), B_z(t) \rangle_T$, where the angle brackets denote an average of the respective magnetic field components over the full duration of the interval $T$. 

\begin{table*}
    \centering
 \begin{tabular}{|c | c | c | c | c | c | c | c | c | c|} 
 \hline
        &           & Interval  & Interval   &           &           &       & Energization Rate  &   Cascade Rate         & \\ 
 Interval & Date      & Start     &  Duration      & $\beta_e$ & $\beta_i$ & Sig.  & $(\partial W_e/\partial t)_{E_\parallel}$        & $\epsilon$  &Ratio\\ 
        &           &           &  $T$ (s)         &           &           &       & ($\times 10^{-12}$~W m$^{-3}$)     & ($\times 10^{-12}$~W m$^{-3}$) &\\ 
 \hline
 00 & 2015-10-16 & 09:24:11 & 70 & 0.08 & 0.83 & S & 5.8 & 2.6 & 2.2\\
 02 & 2016-01-12 & 07:24:38 & 77 & 0.05 & 0.75 & A & 1.7 & 23 & 0.075\\
 03 & 2016-01-13 & 04:56:40 & 27 & 0.13 & 	1.12 & 	A & 	0.42 & 	43 & 	0.001\\
04 & 2016-01-14 & 05:30:26 & 24 & 0.15 & 	1.14 & 	A & 	3.6 & 	33 & 	0.11\\
05 & 2016-01-15 & 02:29:52 & 78 & 0.62 & 	4.26 & 	S & 	1.5 & 	4.4 & 	0.33\\
06 & 2016-01-16 & 07:12:20 & 40 & 0.14 & 	1.11 & 	A & 	14 & 	30 & 	0.48\\
07 & 2016-01-17 & 01:46:28 & 15 & 0.11 & 	0.64 & 	A & 	1 & 	34 & 	0.031\\
08 & 2016-01-18 & 00:50:55 & 62 & 0.17 & 	0.87 & 	A & 	17 & 	16 & 	1.1\\
09 & 2016-01-19 & 07:24:20 & 20 & 0.85 & 	5.21 & 	N & 	1.9 & 	31 & 	0.06\\
10 & 2016-01-20 & 02:52:00 & 150 & 0.32 & 	0.99 & 	A & 	0.65 & 	28 & 	0.023\\
11 & 2016-01-21 & 03:51:45 & 32 & 0.6 & 	2.1 & 	A & 	0.79 & 	15 & 	0.051\\
12 & 2016-01-22 & 00:19:15 & 39 & 1.2 & 	8.67 & 	A & 	1.5 & 	9.3 & 	0.16\\
13 & 2016-01-23 & 02:08:20 & 82 & 0.77 & 	4.19 & 	A & 	2.5 & 	33 & 	0.077\\
14 & 2016-01-24 & 04:29:54 & 49 & 0.25 & 	1.69 & 	A & 	1.3 & 	5.1 & 	0.26\\
15 & 2016-01-25 & 02:18:00 & 20 & 0.45 & 	2.38 & 	A & 	23 & 	28 & 	0.82\\
16 & 2016-01-26 & 00:34:25 & 20 & 0.21 & 	1.41 & 	S & 	1.8 & 	8.1 & 	0.23\\
17 & 2016-01-27 & 06:26:00 & 60 & 0.04 & 	0.34 & 	S & 	7.6 & 	9 & 	0.85\\
18 & 2016-01-28 & 06:05:15 & 32 & 0.1 & 	0.63 & 	S & 	52 & 	350 & 	0.15\\
19 & 2016-01-29 & 23:55:15 & 25 & 0.16 & 	1.55 & 	S & 	14 & 	49 & 	0.28\\
20 & 2016-01-30 & 03:33:00 & 70 & 0.04 & 	0.4 & 	A & 	0.12 & 	1.4 & 	0.09\\
 \hline
 \end{tabular}
    \caption{\label{tab:sample_details} 
    Table of analyzed MMS intervals (all from MMS 1, except for interval 00 which is from MMS 3), including electron energization rates $(\partial W_e/\partial t)_{E_\parallel}$  from the integrated experimental measurements and the theoretically estimated turbulent cascade rate $\epsilon$. The species plasma beta is defined by $\beta_s = \mu_0 n_s \kappa T_s/B^2$. Qualitative signatures are symmetric (S), asymmetric (A), or no clear signature (N). The Ratio column gives $(\partial W_e/\partial t)_{E_\parallel}/\epsilon$.}
\end{table*}

\subsubsection{Electric Field Measurements}
The electric field data is given in GSE coordinates and sampled at 8192 Hz by the Electric Field Double Probes (EDP) instrument suite \blue{\cite{Ergun:2016, Lindqvist:2016, Torbert:2016}}. The electric field measurements in the spacecraft frame, $\V{E'}$(t), are Lorentz transformed \cite{Chen:2011a,Howes:2014} to the mean electron bulk flow frame, \textbf{E}(t) $=\V{E'}(t) + \V{U}_{0e} \times \V{B}$(t) where $\V{U}_{0e}$ is the mean electron bulk flow given by $\V{U}_{0e} = \langle \V{U}_e(t) \rangle_T$.  The $\V{E}$(t) is down-sampled (by averaging) to 30 ms to match the cadence of FPI Dual Electron Spectrometer (DES) \cite{Pollock:2016} and projected onto \textbf{B}$_0$ to obtain $E_\parallel(t)= \V{E}(t) \cdot (\V{B}_0$/$\lvert\V{B}_0\rvert$). Though this downsampling procedure may average over large amplitude, small time scale $E_\parallel$, \newblue{the physical effects of the full electric field on the electrons are implicitly included in the measured electron velocity distributions. This downsampled value is an accurate representation of the electric field at the 30 ms accumulation time of the FPI DES (see \ref{appendix:e_field_analysis}).}

\subsubsection{Electron Velocity Distribution Measurements}
The electron velocity distribution data is sampled at a 30 ms cadence by the FPI DES. Velocity measurements are transformed to the frame of the mean electron bulk flow, $\V{U}_{0e} = \langle \V{U}_e(t) \rangle_T$, with corrections applied in the energy bins to account for any acceleration due to the charged spacecraft. The velocity coordinates are then projected onto the two-dimensional (2V) FAC system, ($v_\parallel$, $v_\bot$), where we integrate over the azimuthal angle in the plane perpendicular to \textbf{B}$_0$ to obtain one perpendicular component ($\bot$) of the electron velocity distribution.

\subsubsection{Measuring Secular Energy Transfer}
The net energy transfer, or \emph{secular energy transfer}, from the parallel electric field in KAWs to electrons has been shown to be mediated by electron Landau damping \cite{Chen:2019}. The signature of electron Landau damping is identified in velocity-space through the use of single-point measurements of  $E_\parallel(t)$ and $f_e(v_\parallel, v_\perp,t)$  from MMS and the FPC technique. The FPC technique utilizes the alternative reduced parallel correlation $C'_{E_{\parallel}}( v_\parallel, t, \tau)$ and the reduced parallel correlation $C_{E_{\parallel}}(v_\parallel, t, \tau)$ to show distinct velocity-space signatures of electron Landau damping and to quantify the energy transferred secularly to the electrons. 

In order to isolate the signature of electron Landau damping, $E_\|(t)$ is high-pass filtered at $f_{cut}=1$~Hz to obtain $\tilde{E}_\parallel(t)$; this serves to remove the low-frequency oscillations associated with undamped wave motion which do not contribute to the net transfer of energy \cite{Howes:2017a}. \blue{The frequency $f_{cut}=1$~Hz is at the lower end of the kinetic range of dissipation, which makes it an appropriate cutoff in order to maintain the higher frequency oscillations that are playing a role in the dissipation. We apply a 5th-order Butterworth filter to the $E_\|(t)$ to obtain the high-pass filtered $\tilde{E}_\parallel(t)$.}

Furthermore, the mean electron velocity distribution $f_{0e}(v_\parallel,v_\bot)= \langle f_e(v_\parallel,v_\bot,t) \rangle_\tau$,
averaged over the interval duration $\tau=T$, is removed from each 30 ms electron distribution to obtain the perturbed electron
velocity distribution, $\delta f_e(v_\parallel,v_\bot,t) = f_e (v_\parallel,v_\bot,t) - f_{0e}(v_\parallel,v_\bot)$. 
In the frame of the mean electron bulk flow, the mean velocity distribution yields zero net energy transfer when 
integrated over velocity, so using the perturbed velocity distribution eliminates a large signal that may obscure 
the bipolar signatures of electron Landau damping. Employing the filtered $\tilde{E}_\parallel(t)$ and perturbed  $\delta f_e(v_\parallel,v_\bot,t)$ to compute the alternative correlation $C'_{E_{\parallel}}(v_\parallel,v_\bot,t, \tau)$ (as in Equation \eqref{eq:full_cprime}) 
and subsequently the correlation $C_{E_{\parallel}}(v_\parallel,t,\tau)$ (as in Equation \eqref{eq:reduced_full_C}) facilitates the identification of the velocity-space signatures of electron Landau damping, as demonstrated in \citeA{Chen:2019}, and as is presented below in the Results section.

\section{Results}

\subsection{Electron Landau Damping and Parallel Poynting Flux Observations}
\begin{figure}
\begin{center}
    \resizebox{6.5in}{!}{\includegraphics[width=\textwidth]{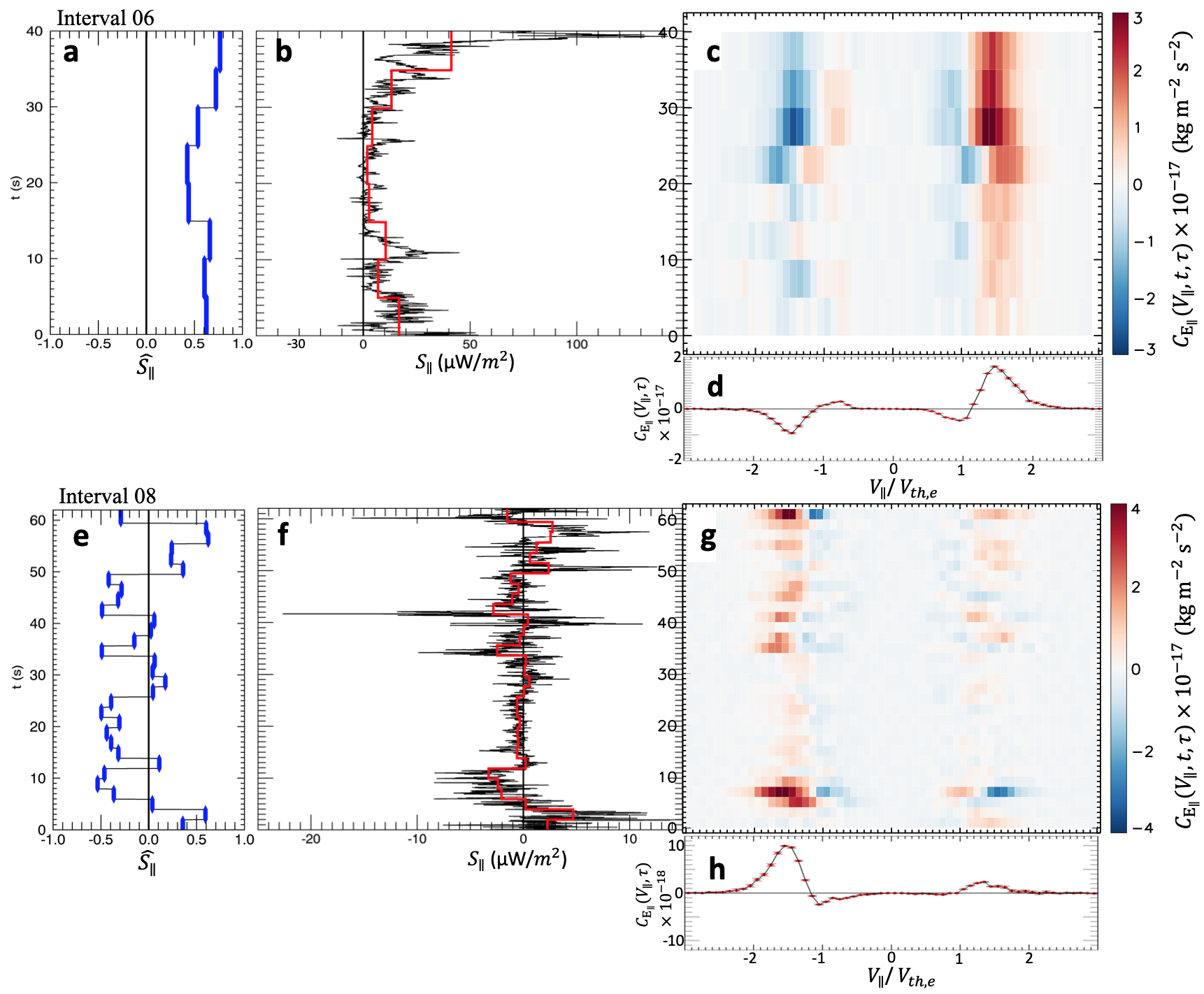}}
\end{center}    
    \caption{\label{fig:poynting_flux}Panels (a-d) are for interval 06 with subintervals $\tau\simeq 5$~s and (e-h) are for interval 08 with subintervals $\tau\simeq 2$~s. The local imbalance of turbulence can be seen in (a,e) the normalized parallel Poynting flux $\hat{S}_\|(t)$ and (b,f) show the amplitude of the parallel Poynting flux $S_\|(t)$. The dissipation of KAWs via electron Landau damping is dominant in the direction of the wave energy flux, as can be seen in (c,g) the timestack plot of $C_{E_{\parallel}}(v_\parallel,t,\tau)$ which yields an asymmetric bipolar signature (d,h) when averaged over the duration of the interval $\tau = T$ to obtain the time-averaged correlation $C_{E_{\parallel}}(v_\parallel,\tau)$.} 
\end{figure}

We present two detailed examples for the identification of electron Landau damping signatures using the timestack plots of the parallel field-particle correlation $C_{E_{\parallel}}(v_\parallel,t,\tau)$, and the comparison of the wave flux directionality with the asymmetric signature of electron Landau damping. We first evaluate the evolution of the electron energization with timestack plots of the parallel field-particle correlation $C_{E_{\parallel}}(v_\parallel,t,\tau)$ over the interval duration in \figref{fig:poynting_flux}(c) for interval 06 with $\tau=4.98$~s and (g) for interval 08 with $\tau=1.98$~s. The time-averaged parallel correlations, taken over the full interval duration $\tau=T$, for these two intervals are plotted in \figref{fig:poynting_flux}(d) and (h). The timestack plots show that while the phase-space energy density transfer rate as a function of $v_\parallel$ varies in time, when its amplitude is significant, it generally has the bipolar form characteristic of the Landau resonance about a parallel velocity, with $v_\parallel/v_{th,e} \sim +1$ for interval 06 (\figref{fig:poynting_flux}(c)) and with  $v_\parallel/v_{th,e} \sim -1$ for interval 08 (\figref{fig:poynting_flux}(g)).  Furthermore, the more clear bipolar velocity-space signatures in the timestack plots are generally found to be persistent in time, indicating ongoing electron Landau damping throughout the interval. \blue{Note that if the electron energization were dominated by energization in the vicinity of coherent structures, such as current sheets, one would expect that the signal of energization would persist only during the short interval while the spacecraft passes through that structure---the persistence in time of these signatures appears to rule out dominance by dissipation mechanisms associated with coherent structures.} \newblue{The lack of current sheets in the intervals analyzed here, because we have chosen intervals with a relatively constant magnetic field, does in fact rule out the possibility that dissipation in current sheets could be responsible for the electron energization signatures observed.}

Linear kinetic theory predicts that, under typical magnetosheath plasma parameters, as the turbulent cascade transfers energy to higher perpendicular wavenumber $k_\perp$, the collisionless damping by electrons becomes strong ($-\gamma/\omega \gtrsim 0.1$) at a point where the resonant parallel phase velocities of KAWs generally reach the electron thermal velocity $v_{th,e}$, satisfying $\omega/k_\parallel v_{th,e} = v_\parallel/v_{th,e} \sim \pm 1$ (see \ref{appendix:collisionless_damping_rates} for predictions of collisionless damping rates of KAWs).  Therefore, the velocities of the zero crossings in the time-averaged, parallel velocity-space signatures observed in \figref{fig:poynting_flux} correspond quantitatively to the resonant velocities expected for significant electron Landau damping. 

\blue{One may question whether a similar bipolar signature at $v_\parallel/v_{th,e} \sim 1$ could be explained the by work done by the parallel electric field $E_\parallel$ on an electron parallel bulk flow $U_{e,\|}$.  A simple calculation indicates that a shifted Maxwellian with a parallel flow of order $U_{e,\|}/v_{th,e} \sim 0.5$ would lead to an energization  signature with a zero crossing near $v_\parallel/v_{th,e} \sim 1$. However, in the mean electron bulk flow frame, $\V{U}_{0e}$, in which our FPC analysis is performed, all of the intervals have $U_{e,\|} \ll v_{th,e}$, ruling out the possibility that these signatures can be explained by a parallel bulk flow correlated with the parallel electric field.  Furthermore, the six symmetric bipolar signatures that we observe cannot be explained by a parallel bulk flow, but can be naturally explained by balanced turbulence, with approximately equal wave energy fluxes up and down the mean magnetic field.}

Also in \figref{fig:poynting_flux}, we present two examples of the parallel Poynting flux for the same two intervals, and compare them with the associated asymmetric bipolar signatures of electron Landau damping for each. We show \figref{fig:poynting_flux}(a) the normalized parallel Poynting flux $\hat{S}_\|(t)$, and (b) the parallel Poynting flux $S_\|(t)$, compared to (c) the timestack plot of $C_{E_{\parallel}}(v_\parallel,t,\tau)$, for interval 06. The normalized parallel Poynting flux and parallel Poynting flux are both positive over the entire interval, so this interval of turbulence is \emph{imbalanced}, dominated by KAWs propagating up the mean magnetic field $\V{B}_0$.  Consistent with this imbalance, the timestack plot of the parallel field-particle correlation $C_{E_{\parallel}}(v_\parallel,t,\tau)$ shows a clear  bipolar, velocity-space signature of electron Landau damping in the $v_\|>0$  direction, persistent in time over the duration of the interval.  Thus, the asymmetry of the velocity-space signature with respect to $v_\parallel = 0$ apparent in (d) the time-averaged correlation $C_{E_{\parallel}}(v_\parallel,\tau)$ is consistent with \emph{locally imbalanced turbulence}, in which electron energization occurs for electrons travelling in the same direction as the dominant \Alfven wave energy flux.

In \figref{fig:poynting_flux}, panels (e-g) we plot the same quantities as described above, but for interval 08. In this interval, the parallel Poynting flux varies about zero, with slightly more wave energy flux, on average, down the mean magnetic field, in the $v_\|<0$ direction.  During some shorter intervals of time, for example $6 \mbox{ s} \le t \le 12 \mbox{ s}$, when the Poynting flux is dominantly down the magnetic field, one sees a clear bipolar signature near $v_\parallel/v_{th,e} \simeq -1$, consistent with energization of those electrons via Landau damping of the KAWs propagating down the magnetic field.  During other periods of time, for example $12 \mbox{ s} \le t \le 25 \mbox{ s}$, one sees that the amplitude of $S_\|(t)$ is small compared to the rest of the interval, and this corresponds to the smaller values of electron Landau damping as seen in panel (g) during the same time. Thus, while there can be variations in the direction of the \Alfven wave energy flux, an asymmetric bipolar signature of electron Landau damping with respect to $v_\parallel = 0$ strongly corresponds with the general direction of the normalized Poynting flux $\hat{S}_\|(t)$ with respect to $\V{B}_0$, and with the amplitude of the wave energy flux, $S_\|(t)$.

\begin{figure}
\begin{center}
    \resizebox{6.5in}{!}{\includegraphics[width=\textwidth]{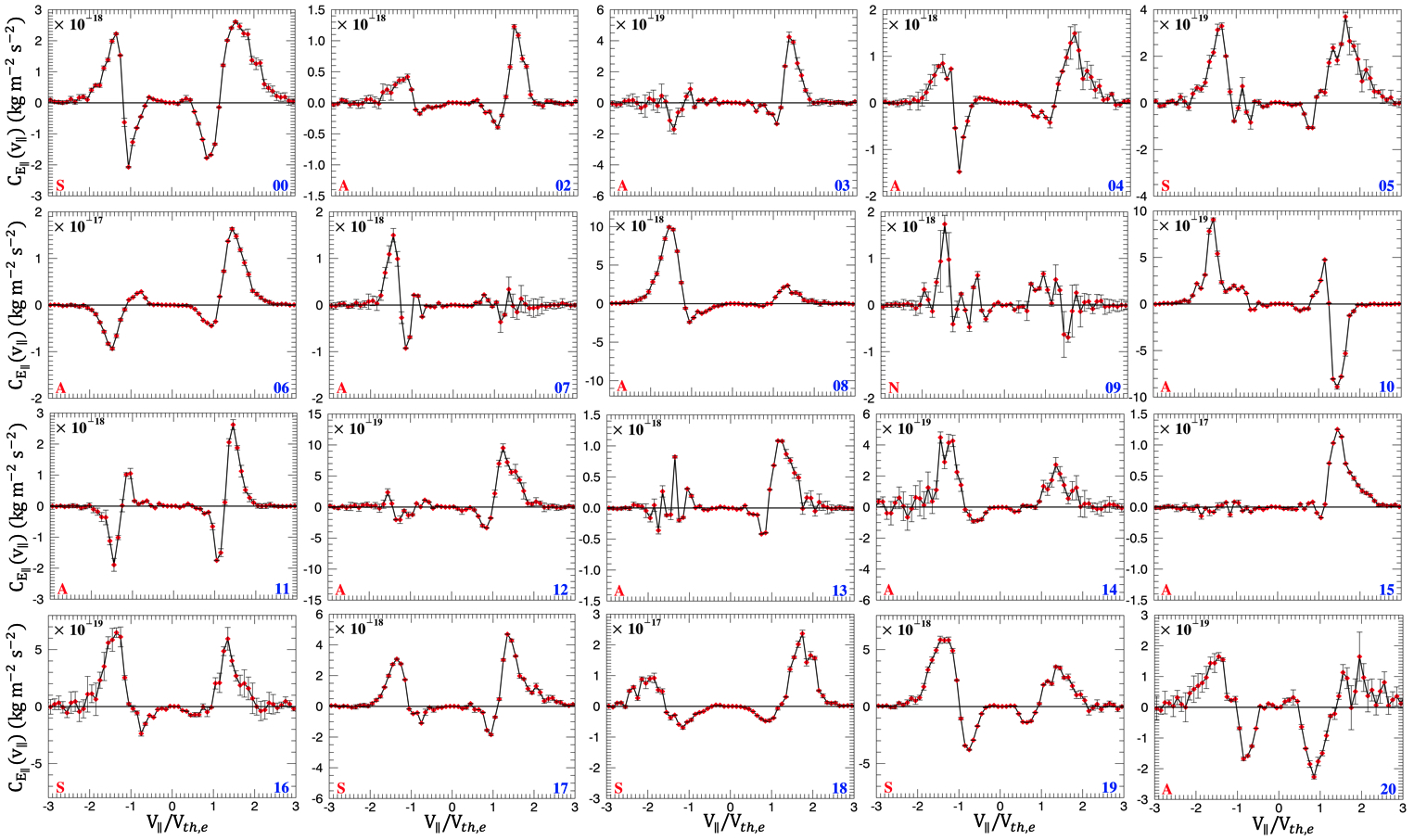}}
\end{center}    
    \caption{\label{fig:all_c_avg_plots} Time-averaged, field-particle correlations $C_{E_{\parallel}}(v_\parallel,\tau)$ for all twenty intervals with $\tau=T$. Qualitative signatures are symmetric (S), asymmetric (A), or no clear signature (N), shown in red at the bottom left of each panel. Interval number is shown in blue at the bottom right of each panel.} 
\end{figure}

The twenty intervals listed in \tabref{tab:sample_details}, including interval 00 from \citeA{Chen:2019}, were analyzed to seek the characteristic bipolar velocity-space signatures of electron Landau damping and their evolution over time by varying the correlation interval $\tau$. In \figref{fig:all_c_avg_plots}, the time-averaged, reduced parallel correlations $C_{E_{\parallel}}(v_\parallel,\tau)$ for all twenty intervals with $\tau=T$ are plotted. Five of the intervals (05, 16, 17, 18, 19) show a velocity-space signature that is approximately symmetric about $v_\parallel = 0$, qualitatively similar to interval 00, but it appears that this result is not typical of all intervals. However, all plots in \figref{fig:all_c_avg_plots}, except interval 09, show a clear bipolar velocity-space signature at either $v_\parallel/v_{th,e} \sim +1$, $v_\parallel/v_{th,e} \sim -1$, or both.  Landau resonant damping of KAWs energizes particles moving approximately at the wave phase velocity \emph{in the direction of wave propagation} resulting in an asymmetric signature (see \figref{fig:Landaudamping_FPC}(c) for an illustration). A recent numerical simulation of the \citeA{Chen:2019} interval has found that the velocity-space signatures of electron Landau damping can have a variety of qualitative appearances \cite{Horvath:2020}, similar to what is observed in \figref{fig:all_c_avg_plots}. In summary, it is found that 19 of the 20 intervals, representing 95\% of the cases, show clear evidence that electron Landau damping is a ubiquitous mechanism for dissipating energy from the turbulent fluctuations in the magnetosheath.  

For all twenty intervals, we calculated the normalized parallel Poynting flux and the parallel Poynting flux according to Equations \eqref{eq:s_para_norm} and \eqref{eq:s_para}, respectively, and find that the wave flux directionality is strongly correlated with the qualitative signature of electron Landau damping. In the six intervals with symmetric signatures (00, 05, 16, 17, 18, 19), the calculated $\hat{S}_\|(t)$ is generally evenly distributed about $\hat{S}_\|(t)=$ 0. The thirteen remaining intervals (excluding 09) exhibit an \emph{imbalance} of wave energy flux with $S_\|(t) < $ 0 for anti-parallel wave flux, and $S_\|(t) >$ 0 for parallel wave flux, corresponding to the asymmetric bipolar signatures seen in \figref{fig:all_c_avg_plots}. Therefore, we interpret the thirteen intervals of asymmetric velocity-space signatures in \figref{fig:all_c_avg_plots} to indicate \emph{locally imbalanced turbulence}. The qualitative signatures (symmetric, asymmetric, or no clear signature) of electron Landau damping for all twenty intervals are listed in \tabref{tab:sample_details}.

\subsection{Comparison of Turbulent Energy Dissipation to Turbulent Energy Cascade Rate}
\begin{figure}[htbp]
\begin{center}
  \resizebox{5.0in}{!}{\includegraphics{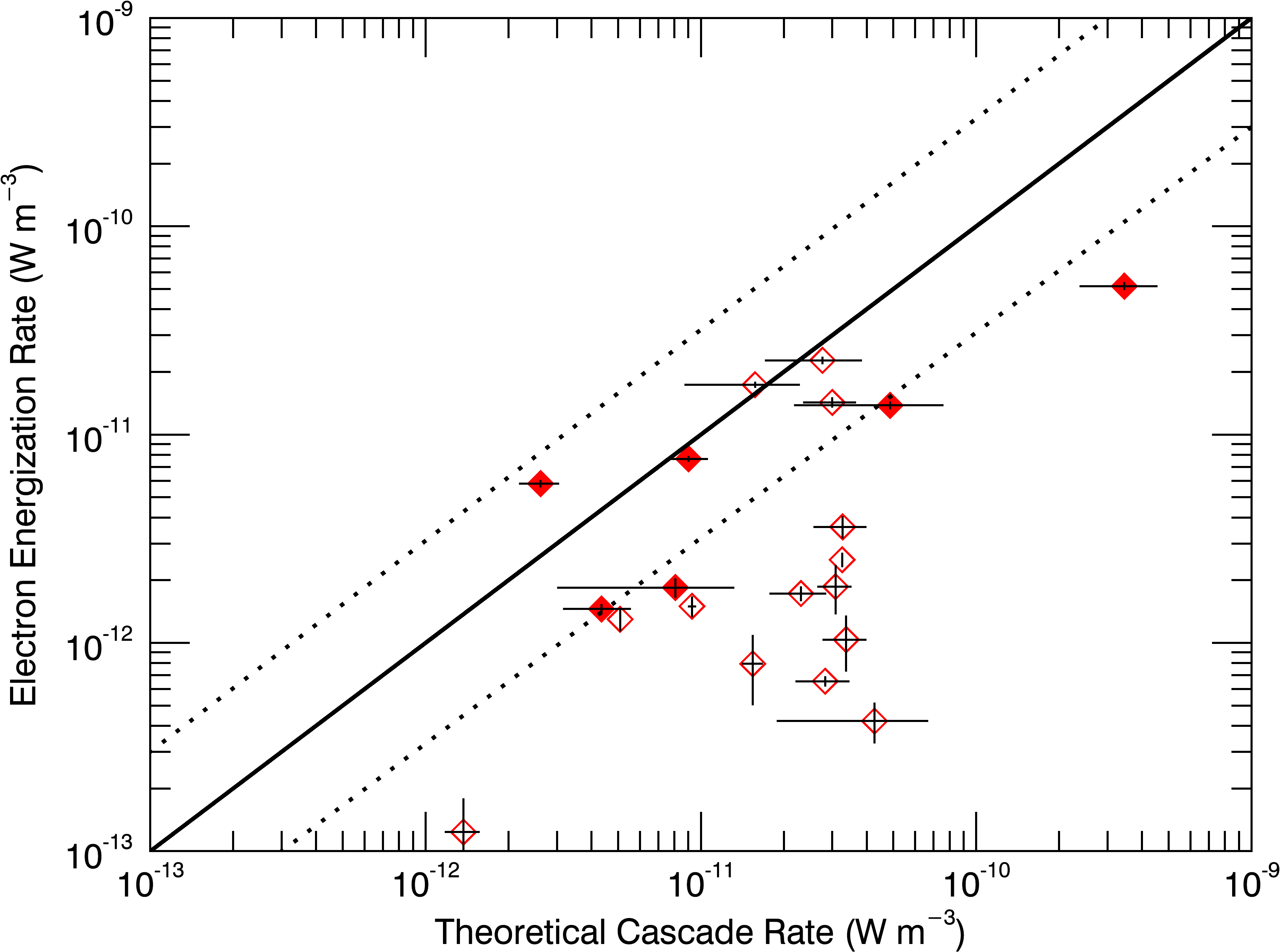} } 
\end{center}
    \caption{\label{fig:cascade} The electron energization rate $(\partial W_e/\partial t)_{E_\parallel}$ versus the theoretical cascade rate $\epsilon$. The solid line indicates $(\partial W_e/\partial t)_{E_\parallel}=\epsilon$ and the dotted lines indicate the range of the order-of-magnitude estimate of $\epsilon$. Solid diamonds represent symmetric signatures, and open diamonds represent asymmetric signatures.}
\end{figure}

Having established the ubiquity of electron Landau damping in magnetosheath turbulence, and having shown that asymmetric bipolar signatures are consistent with intervals of locally imbalanced turbulence, we next determine the fraction of the turbulent dissipation governed by electron Landau damping. Using Equation \eqref{eq:eenergization}, we have calculated the electron energization rates for all twenty intervals; they are listed in  \tabref{tab:sample_details}.  Note that, although the reversible nature of collisionless energy transfer allows for negative values that indicate the transfer of energy from electrons to $E_\parallel$, in all twenty intervals a net transfer of energy to the electrons is found.  For each interval, the electron energization rate is compared with a theoretically estimated turbulent energy cascade rate $\epsilon$ given by Equation \eqref{eq:eps}. The theoretical cascade rates $\epsilon$, listed in \tabref{tab:sample_details}, are plotted against the measured electron energization rates $(\partial W_e(\V{r}_0,t)/\partial t)_{E_\parallel}$ in \figref{fig:cascade}.

The solid line in \figref{fig:cascade} indicates $(\partial W_e/\partial t)_{E_\parallel}= \epsilon$, meaning that the measured rate of energization by electron Landau damping is sufficient to remove all of the energy from the turbulent cascade.  However, the cascade rate estimate, derived from a scaling theory, is an \emph{order-of-magnitude} calculation, so dotted lines at a factor of 3 above and below indicate the range of accuracy of this order-of-magnitude estimate for $\epsilon$.  Our key result is that \blue{roughly one third} of the intervals have measured electron energization rates that fall within the \emph{order-of-magnitude estimate} for $\epsilon$, meaning that electron Landau damping is frequently a major mechanism responsible for the dissipation of turbulent energy in these magnetosheath intervals. 

\section{Discussion}

\subsection{Connection to Observational Determinations of Turbulent Cascade and Dissipation Rates}
Modern spacecraft instrumentation on missions such as Cluster \cite{Escoubet:1997}, THEMIS \cite{Angelopoulos:2008}, and MMS \cite{Burch:2016} enables the turbulent cascade and dissipation rates to be determined observationally using several different approaches.  Various forms of Komolgorov's scaling law for third-order structure functions modified for magnetized plasma turbulence \cite{Politano:1998a, Banerjee:2013, Andres:2018} (so-called third-order laws) utilize a long time series of magnetic field and plasma bulk velocity fluctuation measurements to estimate statistically the turbulent energy cascade rate.  Examining well developed turbulence in the Earth's magnetosheath, recent studies using different formulations of third-order laws have found turbulent cascade rates of  $\epsilon \in [10^{-16},10^{-13}]$ W m$^{-3}$ using 47 intervals of  \Alfvenic turbulence from Cluster and THEMIS \cite{Hadid:2018}, $\epsilon \in [10^{-13},10^{-12}]$ W m$^{-3}$ using three MMS intervals \cite{Bandyopadhyay:2018Oct, Bandyopadhyay:2020June, Bandyopadhyay:2020June_b}, and $\epsilon \in [10^{-13},10^{-12}]$~W~m$^{-3}$ using 72 MMS intervals \cite{Andres:2019}.   

A complementary approach is to measure the particle energization rates by computing directly the work done on the plasma by the electric field through  $\V{J} \cdot \V{E}$.  For a single MMS interval of magnetosheath turbulence with significant ion cyclotron wave power, a time-averaged rate of electromagnetic work $\langle \V{J} \cdot \V{E} \rangle \sim 10^{-14}$~W m$^{-3}$ was found \cite{He:2019}.  Alternatively, the conversion of turbulent energy into internal energy by pressure-strain interactions was determined using the ``Pi-D'' method for a single MMS interval, obtaining  $\langle \mbox{Pi-D}^p \rangle \sim 5 \times 10^{-13}$~W~m$^{-3}$ for protons and $\langle \mbox{Pi-D}^e \rangle \sim 4 \times 10^{-13}$~W~m$^{-3}$ for electrons \cite{Bandyopadhyay:2020June_b}.  It is worth noting, however, that the third-order law cascade rate estimate for the MMS interval in this latter study yielded $\epsilon \sim 10^{-13}$~W~m$^{-3}$, a full order of magnitude lower than the summed proton and electron energization rates.

It is clear that the turbulent cascade rates estimated by the third-order laws fall at the low end of, or even well below, the electron energization rates measured herein. In addition, the electromagnetic work and pressure-strain rates fall at the lower end of or slightly below our measured range for the electron energization rates.  It is worthwhile emphasizing that the electron energization rates computed herein have been validated by direct comparisons to the interval-averaged $\langle J_{\parallel,e}E_\parallel \rangle_T$ (see \ref{appendix:quality_checks}) and are consistent with the turbulent energy cascade rates predicted by our model \cite{Howes:2008,Howes:2011}, meaning that $(\partial W_e/\partial t)_{E_\parallel} \lesssim \epsilon$.
The origin of the broad disagreement between our results from the twenty MMS intervals and these previous studies is not immediately apparent, but a few comments are germane.  First, dissipation rates can vary dramatically in time (for example, see the timestack plot for interval 08 in \figref{fig:poynting_flux}(g)), so instantaneous rates and statistically averaged rates may differ significantly.  Second, the derivation of third-order laws generally depends on simplifying assumptions --- \emph{i.e.}, that the cascade of energy is isotropic relative to the direction of the magnetic field, an assumption which appears to be at odds with direct multi-spacecraft measurements of the distribution of turbulent power \cite{Sahraoui:2010b, Roberts:2013, Roberts:2015b} --- so it is possible that these magnetosheath intervals do not satisfy the conditions needed to apply these particular formulations of the third-order law. This issue could explain how the direct energization rates computed herein and in \citeA{Bandyopadhyay:2020June_b} are larger than the 3rd-order law cascade rates.  A larger statistical study, and cross comparisons between existing methods, will enable the root of the disagreement to be discovered.

\subsection{Implications for Models of Collisionless Plasma Turbulence}
Early theoretical investigations of magnetized plasma turbulence in the framework of incompressible MHD emphasized the wave-like nature of turbulent plasma motions, suggesting that nonlinear interactions between counterpropagating \Alfven waves --- or \Alfven wave collisions --- mediate the turbulent cascade of energy from large to small scales \cite{Iroshnikov:1963,Kraichnan:1965}.  These insights motivated studies of  weak incompressible MHD turbulence \cite{Sridhar:1994,Ng:1996,Galtier:2000}, eventually leading to an analytical solution of the nonlinear energy transfer due to \Alfven wave collisions in the weakly nonlinear limit \cite{Howes:2013a} that has been confirmed numerically \cite{Nielson:2013a} and verified experimentally in the laboratory \cite{Howes:2012b}.  In the strongly nonlinear limit, the nonlinear physics of \Alfven waves remains central to modern theories of MHD turbulence, providing the physical foundation for explaining the anisotropic nature of the turbulent cascade \cite{Goldreich:1995}, the dynamic alignment of magnetic field and velocity fluctuations \cite{Boldyrev:2006}, and the self-consistent generation of current sheets in plasma turbulence \cite{Howes:2016b,Verniero:2018a,Verniero:2018b}. Extending the physics of plasma turbulence into the weakly collisional limit, the linear and nonlinear wave physics continues to provide a valuable theoretical framework for the construction of models of the turbulent cascade and its dissipation  \cite{Howes:2008,Schekochihin:2009,Howes:2015b}. 

On the other hand, alternative models of plasma turbulence reject the importance of linear wave physics to the nonlinear evolution \cite{Matthaeus:2014}, focusing instead on the development of coherent structures, specifically current sheets, in which the dissipation of turbulent energy is found to be largely concentrated \cite{Uritsky:2010,Osman:2011,Zhdankin:2013}. In the quest to identify the mechanisms that remove energy from the turbulent fluctuations in collisionless plasma turbulence, this question of the relative importance of waves versus coherent structures in turbulence has defined an important frontier in space physics and astrophysics  \cite{Groselj:2019}.  A recent review suggested that ``coherent structures and associated non-uniform dissipation play a very important and possibly dominant role in the termination of the cascade and the effectively irreversible conversion of fluid macroscopic energy into microscopic random motion'' \cite{Matthaeus:2015}. 

Since collisionless damping via the Landau resonance is inherently linked to the physics of waves through resonant interactions at the wave phase velocity, our observational finding that electron Landau damping plays a significant role in the collisionless removal of energy from turbulent fluctuations, and in particular the Poynting flux analysis identifying intervals of locally imbalanced turbulence, directly contradicts the claim that coherent structures, and not waves, dominate the physics of turbulent dissipation in space and astrophysical plasmas. In fact, kinetic numerical simulations of \Alfven wave collisions in the strongly nonlinear limit, defined by critical balance \cite{Goldreich:1995}, have shown that current sheets self-consistently arise from the nonlinear wave physics through constructive interface among the initial finite-amplitude \Alfven wave modes and the nonlinearly generated \Alfven modes \cite{Howes:2016b,Verniero:2018a,Verniero:2018b}. This finding undercuts the attempt to argue that the presence of current sheets is evidence that the strongly interacting \Alfven wave interpretation of plasma turbulence is somehow incomplete \cite{Matthaeus:2011}.  

Furthermore, \blue{although it has often been suggested} that magnetic reconnection plays a role in the dissipation 
of plasma turbulence \cite{Osman:2011,Karimabadi:2013a,Loureiro:2017a,Mallet:2017a}, there remains a lack of observational evidence for magnetic reconnection playing a role in the dissipation of plasma turbulence in any space environment, such as Earth's magnetosheath. In fact, a recent FPC analysis of the ion and electron energization in a kinetic simulation of the current sheets arising from strong \Alfven wave collisions has shown that Landau damping dominates the spatially non-uniform particle energization in the vicinity of those current sheets \cite{Howes:2018a}, a result demonstrating that the observation of particle energization in the vicinity of current sheets in the solar wind \cite{Osman:2011, Osman:2012a, Osman:2012b,wu:2013a, Osman:2014b} is not necessarily evidence of magnetic reconnection.  Although magnetic reconnection requires the presence of current sheets, the presence of current sheets alone does not imply that magnetic reconnection is occuring.

Earlier work has suggested that the nonlinear interactions in strong turbulence could lead to the inhibition of Landau damping as an effective means for removing energy from the turbulent fluctuations \cite{Plunk:2013}, but our observational findings clearly settle this question: Landau damping can indeed play an effective role in dissipating strong plasma turbulence. The effectiveness of Landau damping has been further called into question with the recent discovery that, in a sufficiently collisionless environment, such as the solar wind, the Landau damping of compressible fluctuations within the inertial range (at scales $k_\perp \rho_i \ll 1$) is entirely suppressed by a nonlinearly driven stochastic plasma echo, which cancels the effect of linear phase mixing and thereby leads to a ``fluidization'' of the compressive fluctuations at those scales \cite{Meyrand:2019}.  Our results clearly prove that electron Landau damping of turbulent KAW fluctuations, which occurs at sub-ion scales $k_\perp \rho_i \gg 1$, can lead to a removal of energy from the turbulence and net energization of the electrons. The nonlinear kinetic physics underlying the change in the effectiveness of Landau damping at these different length scales remains to be elucidated.

\subsection{Implications for Predictive Models of Turbulent Plasma Heating}
The development of a predictive model of the differential heating of the plasma species by turbulence will facilitate improved interpretations of astronomical observations in terms of the unresolved plasma dynamics and enable better predictions of how microphysical plasma heating impacts the evolution of the coupled solar-terrestrial system.  Successful models will require the identification of the physical mechanisms governing the turbulent dissipation and the determination of the their relative contributions to the total dissipation as a function of the plasma and turbulence parameters.  Although the twenty intervals analyzed here are insufficient to draw conclusions about the relative contribution of electron Landau damping as a function of the plasma parameters $\beta_i$ and $\beta_e$,  this study demonstrates that the FPC technique is a viable approach to meet this challenge.  

Furthermore, the FPC technique can be used to identify the unique velocity-space signatures of other proposed collisionless energy transfer mechanisms, including ion Landau damping \cite{Klein:2017}, ion cyclotron damping \cite{Klein:2020}, stochastic ion heating \cite{Cerri:2021}, and magnetic pumping \cite{Montag:2021}. A quantitative determination of the contribution of each of these mechanisms to the turbulent dissipation as a function of the plasma and turbulence parameters using the techniques presented here would present a major achievement in the long-term quest to develop predictive models of how turbulence couples large-scale dynamics to the particle energization and plasma heating that govern the evolution of space and astrophysical plasma environments.

\section{Conclusion}

We performed a field-particle correlation (FPC) analysis of a sample of 20 intervals of \Alfvenic magnetosheath turbulence measured by the MMS spacecraft.  We have identified the velocity-space signature of electron Landau damping in 95\% of these intervals, with resonant zero crossings of the bipolar signatures that are quantitatively consistent with the theoretical expectation for KAWs that the resonant parallel phase velocities have values $\omega/k_\parallel v_{th,e} \simeq \pm 1$ as the collisionless damping becomes strong.  These findings confirm the first definitive identification of electron Landau damping in the magnetosheath by  \citeA{Chen:2019}, and provide direct evidence that electron Landau damping is a nearly ubiquitous dissipation mechanism in these intervals of \Alfvenic turbulence. However, it is found that only about one quarter of the intervals produce a symmetric pair of bipolar signatures at $v_\parallel/ v_{th,e} \sim \pm 1$, as seen in \citeA{Chen:2019}. A Poynting flux analysis shows that asymmetric signatures arise when there is more \Alfven wave energy flux in one direction along $\V{B}_0$ than the other over the duration of the observed interval, where  the dissipation of such locally imbalanced turbulence will result in a larger bipolar signature on one side. The observed prevalence of asymmetric signatures indicates that one more often finds locally imbalanced turbulence, a finding consistent with the analysis of a recent kinetic numerical simulation of magnetosheath turbulence \cite{Horvath:2020}.  

Beyond the unequivocal identification of electron Landau damping, we assess its contribution to the total dissipation rate of the turbulence. As shown in \figref{fig:cascade}, by comparing the measured electron energization rates by Landau damping $(\partial W_e/\partial t)_{E_\parallel}$ to the theoretically estimated turbulent energy cascade rates $\epsilon$, our key finding is that electron Landau damping accounts for the dissipation of a significant fraction, if not a majority, of the total turbulent energy in \blue{about one third} of the intervals.  We estimated turbulent energy density cascade rates in the range $\epsilon \in [10^{-12},10^{-10}]$~W~m$^{-3}$ and measured rates of change of electron energy density in the range  $(\partial W_e/\partial t)_{E_\parallel} \in [10^{-13},10^{-11}]$~W~m$^{-3}$. Future work will relax the restriction of the FPC analysis to intervals with relatively constant magnetic field magnitude and density, enabling a larger statistical sample of MMS burst-mode intervals to be analyzed and to rule out selection bias in these results.

\appendix
\section{Quality Checks}
\label{appendix:quality_checks}
\begin{figure}
\begin{center}
    \resizebox{5.0in}{!}{\includegraphics[width=0.3\textwidth]{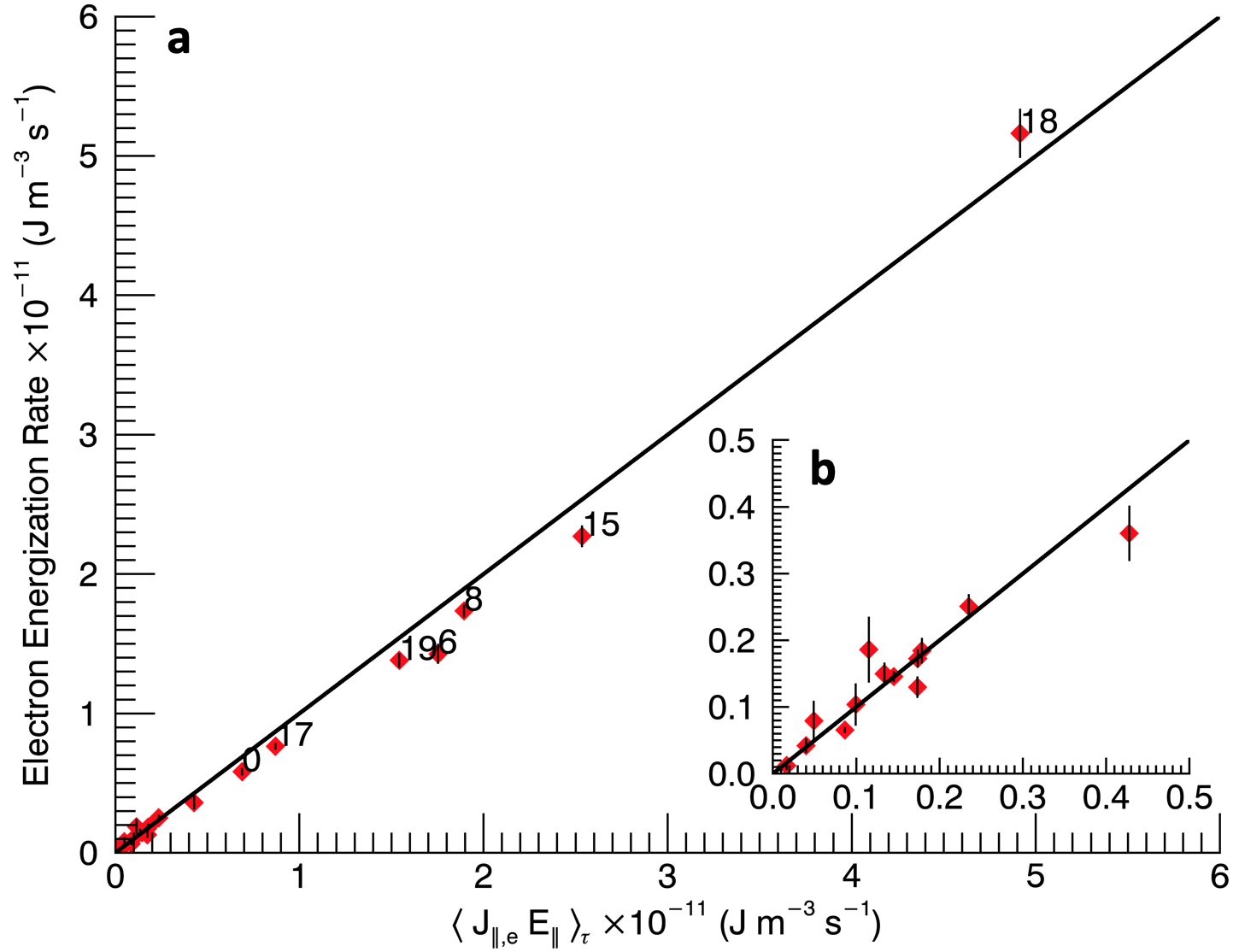}}
\end{center}    
    \caption{\label{fig:jdote} (a) Comparison of $\left(\partial W_e/\partial t\right)_{E_\parallel}$ calculated using the FPC technique, versus $\langle J_{\parallel, e} E_\parallel \rangle_\tau$ calculated using FPI moments for all twenty intervals. The solid line indicates where $\partial W_e/\partial t = \langle J_{\parallel, e} E_\parallel \rangle_\tau$. (b) The inset is zoomed to the lower values. }
\end{figure}

Integrating the correlation $C_{E_{\parallel}}(\V{v}, t, \tau)$ over velocity space and averaging over the full duration of the interval, $\tau=T$, yields the average rate of work done on the electrons by the parallel electric field, equal to the time average of the contribution to $\V{J}_e\cdot\V{E}$ by $E_\parallel$,
\begin{equation}
\left(\frac{\partial W_e}{\partial t}\right)_{E_\parallel} = \int d{\bf v} \ C_{{E_\parallel}}(\V{v},t,\tau) = \langle J_{\parallel,e}(t)E_\parallel(t) \rangle_\tau = \langle -\lvert q_e \rvert n_e U_{\|,e}(t) E_\|(t) \rangle_\tau.
\label{eq:energy_density_transfer_rate}
\end{equation}
Here $q_e$ is the electron charge, $n_e$ is the electron number density, and $U_{\|,e}$ is the component of the electron bulk velocity parallel to \textbf{B}$_0$; the values of the electron number density and electron bulk velocity are from the FPI moments. The $E_\|$ used in the calculation of $\langle J_{\parallel,e}E_\parallel \rangle_\tau$ is the same high-pass filtered $\tilde{E}_\|$ used in the calculation of the correlation $C_{E_{\parallel}}({\bf v}, t, \tau)$. \blue{This $\tilde{E}_\|$ has been shifted to the mean electron bulk flow frame as described in Section 2.4.3.} The comparison of $\left(\partial W_e/\partial t\right)_{E_\parallel}$ with $\langle J_{\parallel,e}E_\parallel \rangle_\tau$ for all twenty intervals is shown in \figref{fig:jdote}. We see that all of the values are close to the one-to-one line, indicating strong agreement between two distinct methods for calculating the electron energization rates. This cross-check validates the electron energization rates obtained using the FPC technique.

\section{Error Bar Calculation}
The uncertainty associated with the calculation of the correlation $C_{E_{\parallel}}(v_\parallel)$ shown in \figref{fig:poynting_flux}(d) and (h), and \figref{fig:all_c_avg_plots}, and for the electron energization rates $\left(\partial W_e/\partial t\right)_{E_\parallel}$ shown in \figref{fig:cascade} and \figref{fig:jdote}, were calculated using the distribution uncertainty reported in the L2 data for FPI DES. The uncertainty reported therein is the Poisson counting statistics error where the error goes as the inverse of the square root of the number of counts, $\sim$N$^{-1/2}$. The uncertainty in each step of the calculation was calculated using the methods described in \citeA{Taylor:1997}, where a constant multiplied by the distribution measurement value will have an associated uncertainty of the same constant multiplied by the uncertainty in that distribution measurement. Anytime averages, differences, or sums are taken, the associated uncertainty is calculated in quadrature; \blue{see \tabref{tab:error_analysis}}. 

\blue{It is important to note that in this propagation of Poisson uncertainty we have assumed the electric field to be a constant, thus having no systematic uncertainty. The errors reported in the Level 2 parallel electric field data from EDP apply to the DC electric field and not to the electric field fluctuations above 1 Hz; the electric fluctuations above 1 Hz are generally reliable. The full propagation of errors $-$  including the systematic errors in the magnetic and electric fields, and in the plasma moments \cite{Gershman:2019}, as well as the Poisson errors in the distributions measurements \cite{Gershman:2015} $-$ is a vastly more complex technique and currently beyond the scope of this work.}

The error bars associated with the theoretical cascade rate $\epsilon$ in \figref{fig:cascade} represent the calculation of Eq. \ref{eq:eps} using $\delta \hat{B}_\perp(f) \pm \sigma$ where $\sigma$ is standard deviation of $\delta \hat{B}_\perp(f)$. The $\pm \sigma$ corresponds to the upper and lower bounds of the uncertainty in $\epsilon$, respectively.

\begin{table*}
    \centering
 \begin{tabular}{|c | c | c | c |} 
 \hline
        & symbol & associated uncertainty & description          \\ 
 \hline
 original data & $f_e (v_\parallel,v_\bot,t)$ & $\sigma = \frac{f_e (v_\parallel,v_\bot,t)}{\sqrt{N}}$ & N is the number of counts; \\
 &  &   & given in FPI DES Level 2 data. \\
 \hline
 background distribution & $f_{0e}(v_\parallel,v_\bot)$ & $\sigma' = \frac{1}{n} \sqrt{\sum\limits_j^n  \sigma_j^2}$ & the sum is over time at each $v_\| - v_\bot$ \\
 &  &   & coordinate; n $=\tau$/0.03. \\
 \hline
 fluctuations & $\delta f_e(v_\parallel,v_\bot,t)$ & $\sigma'' = \sqrt{\sigma'^2 + \sigma^2}$ & $\sigma'' = \sigma''(v_\|,v_\bot)$. \\
 \hline
alternative correlation & $C'_{E_{\parallel}}(v_\parallel, v_\perp, t, \tau)$ & $\sigma''' = q_e v_{\|,j} \sigma''(v_{\|,j}, v_\bot) E_\|$ & $\forall$ j in $v_\|-$space; $\sigma''' = \sigma'''(v_\|,v_\bot)$; \\
&   &   & associated with Eq. \eqref{eq:full_cprime}. \\
\hline
alternative correlation binned & $C'_{E_{\parallel}}(v_\parallel, v_\perp, t, \tau)$ & $\sigma^{(4)} = \frac{1}{n} \sqrt{\sum\limits_j^n \sigma_j^{'''2}}$ & n is the number of measurements \\
&  &   & in each square bin $\Delta v_\bot = \Delta v_\|$. \\
reduced alternative correlation & $C'_{E_{\parallel}}(v_\parallel, t, \tau)$ & $\sigma^{(5)} = 2\pi \Delta v_\bot \sqrt{\sum\limits_j (v_{\bot,j}\sigma^{(4)})^2}$ & the sum is over $v_\bot$ from 0 to 3$*v_{th,e}$; \\
&   &   & associated with Eq. \eqref{eq:reduced_cprime}. \\
\hline
reduced correlation & $C_{E_{\parallel}}(v_\parallel, t, \tau)$ & $\sigma^{(6)} = \frac{v_{\|,j}}{2\Delta v_\|} \sigma^{(5)}(v_{\|,j})$ & $\forall$ j in $v_\| - $space; $\sigma^{(6)} = \sigma^{(6)}(v_\|)$; \\
&   &   & associated with Eq. \eqref{eq:reduced_full_C}. \\
\hline
time averaged correlation & $C_{E_{\parallel}}(v_\parallel, t = \tau)$ & $\sigma^{(7)} = \frac{1}{n} \sqrt{\sum\limits_j^n \sigma_j^{(6)2}}$ & the sum is over time at each $v_\|$  \\
&  &   & coordinate; n $=\tau$/0.03. \\
\hline
energy density transfer rate & $\left(\frac{\partial W_e(\V{r}_0,t)}{\partial t}\right)_{E_\parallel}$ & $\sigma_f = \Delta v_\| \sqrt{\sum_j \sigma_j^{(7)2}}$ &  the sum is over $v_\|$ from \\
&   &   & $-$3$v_{th,e}$ to $+$3$v_{th,e}$; \\
&   &   & $\sigma_f$ is the final error reported \\
&  &   & for the electron energization rate; \\
&   &   & associated with Eq. \eqref{eq:eenergization}. \\
 \hline
 \end{tabular}
    \caption{\label{tab:error_analysis} 
    Table of uncertainty analysis steps starting with the Poisson uncertainty associated with the electron distribution measurement given in the FPI DES Level 2 data. The square bin sizes are 10\% the electron thermal speed, $\Delta v_\bot = \Delta v_\| = 0.1v_{th,e}$.}
\end{table*}

\section{Linear Kinetic Theory Predictions of Collisionless Damping Rates}
\label{appendix:collisionless_damping_rates}
\begin{figure}
\begin{center}
    \resizebox{5.0in}{!}{\includegraphics[width=0.3\textwidth]{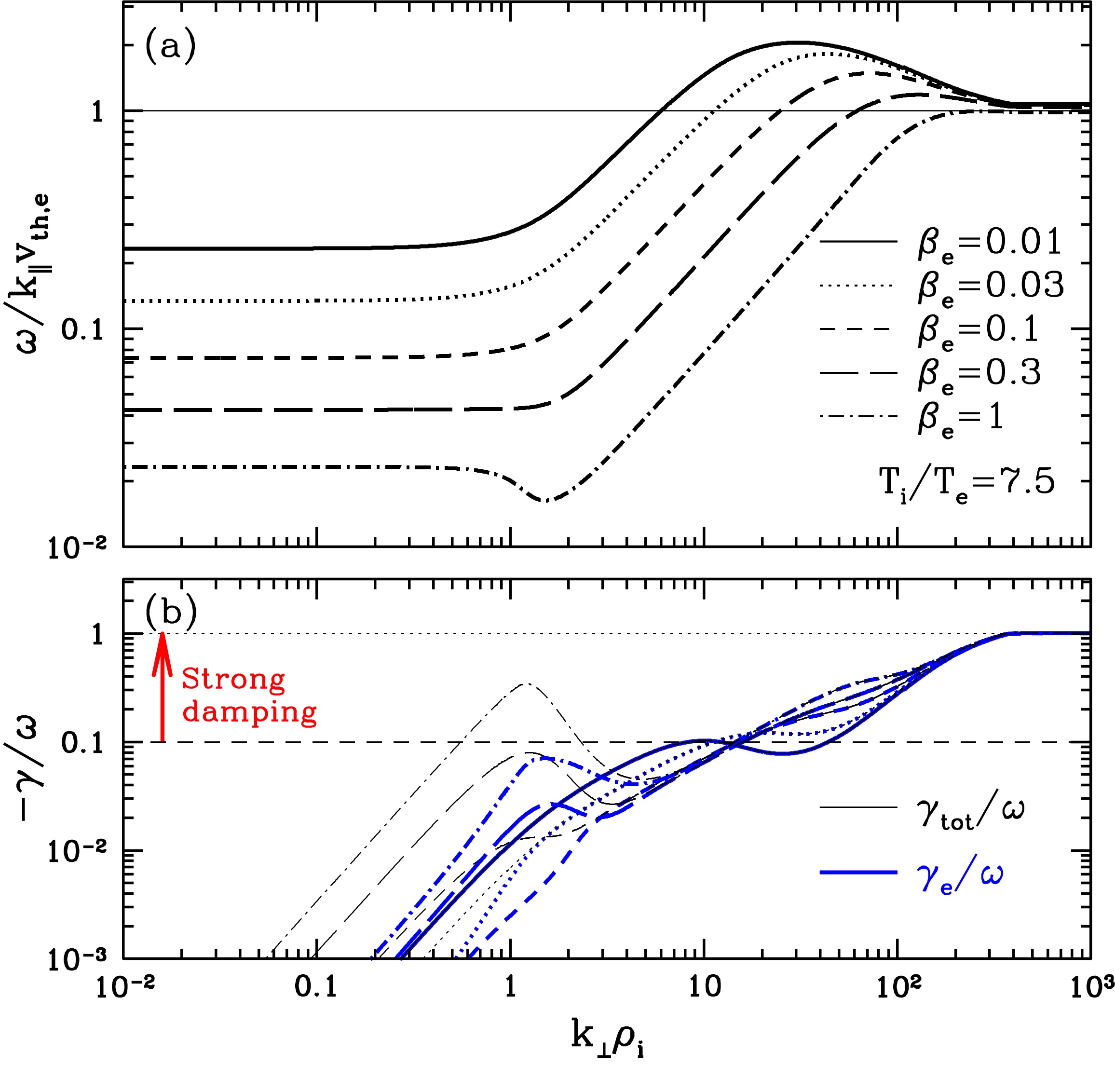}}
\end{center}    
    \caption{\label{fig:disprel} Predictions from the Vlasov-Maxwell linear dispersion relation for the collisionless damping of KAWs in a plasma with parameters $\beta_e \in [0.01,~0.03,~0.1,~0.3,~1]$ and $T_i/T_e=7.5$. Panel (a) shows the parallel phase velocity normalized to the electron thermal velocity $\omega/k_\parallel v_{th,e}$ vs.~the normalized perpendicular wavenumber $k_\perp \rho_i$ and panel (b) shows the total normalized damping rate $\gamma_{tot}/\omega$ (thin black) and damping rate due to electrons only $\gamma_{e}/\omega$ (thick blue) vs.~the normalized perpendicular wavenumber $k_\perp \rho_i$.  Collisionless damping becomes strong for values $-\gamma/\omega \gtrsim 0.1$.}
\end{figure}

Using the Vlasov-Maxwell linear dispersion relation for a fully ionized proton and electron plasma with isotropic Maxwellian equilibrium velocity distributions and a realistic mass ratio $m_p/m_e=1836$, we plot in \figref{fig:disprel}(a) the parallel phase velocity $\omega/k_\parallel$ vs.~$k_\perp \rho_i$, where the ion Larmor radius is $\rho_i=v_{th,i}/\Omega_i$, the ion thermal velocity is $v^2_{th,i}=2\kappa T_i/m_i$, and the ion cyclotron frequency is $\Omega_i=q B/m_i$.   In panel (b), the total normalized collisionless damping rate $-\gamma/\omega$ is plotted vs.~$k_\perp \rho_i$, where we present both the total (ion plus electron) collisonless damping rate $-\gamma_{tot}/\omega$ (thin black curves) and the collisionless damping due to electrons alone $-\gamma_{e}/\omega$ (thick blue curves).  A rule of thumb is that collisionless damping by electrons becomes strong for $-\gamma_{e}/\omega \gtrsim 0.1$ (horizontal dashed black line).  For the plasma parameters, we take the average value of $T_i/T_e=7.5$ from the 20 intervals in \tabref{tab:sample_details}.  For the range of electron plasma beta values $0.01 \le \beta_e \le 1$ in our sample, where $\beta_e = 2 \mu_0 n_e T_e/B^2$, the plots shows that the electron  parallel phase velocity generally reaches the electron thermal velocity, giving $\omega/k_\parallel v_{th,e} \sim 1$, when the collisionless damping by electrons becomes strong, $-\gamma_{e}/\omega \gtrsim 0.1$.  The physical reason for this result is that, when the parallel phase velocity increases to $v_{th,e}$, the negative slope of the electron distribution function peaks, leading to strong collisionless damping by the electrons.
This means that we expect the zero crossing in the bipolar signature of electron Landau damping presented in this paper to appear at resonant parallel velocities $v_\parallel/v_{th,e} \sim \pm 1$, as observed in Figures~\ref{fig:poynting_flux} and~\ref{fig:all_c_avg_plots}.

\section{Electric Field Analysis}
\label{appendix:e_field_analysis}

\begin{figure}
\begin{center}
    \resizebox{5.0in}{!}{\includegraphics[width=0.3\textwidth]{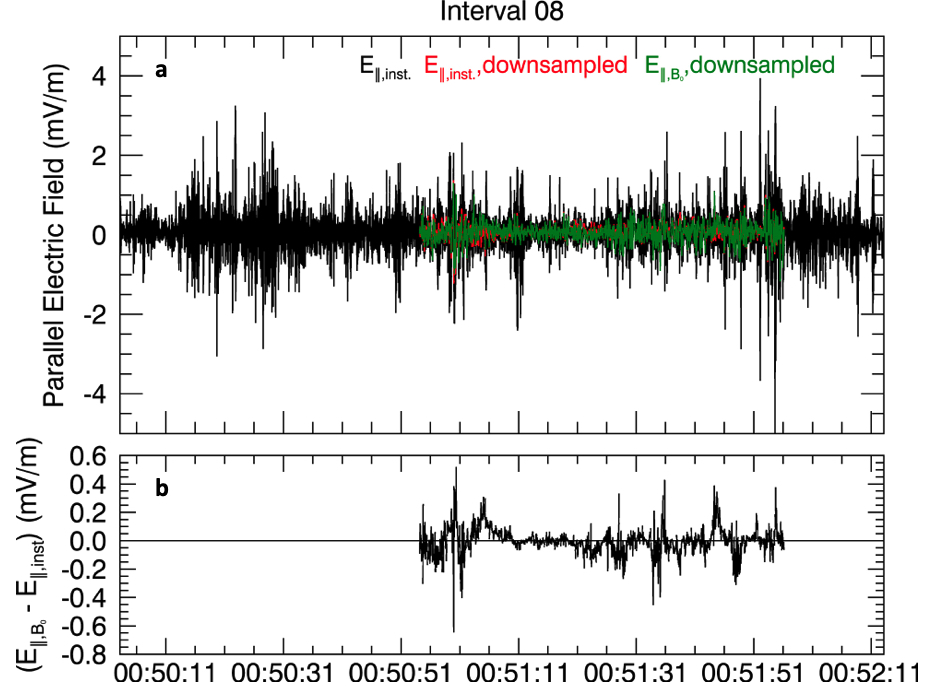}}
\end{center}    
    \caption{\label{fig:interval08_e_field} Parallel electric field measurements for interval 08: (a) the electric field parallel to the instantaneous magnetic field at full EDP cadence ($\V{E}_{\|,inst}$, black), the electric field parallel to the instantaneous magnetic field downsampled to FPI DES cadence ($\V{E}_{\|,inst}$, red), the electric field parallel to the ambient magnetic field downsampled to FPI DES cadence ($\V{E}_{\|,B_0}$, green); (b) the difference between the downsampled $\V{E}_{\|,B_0}$ and $\V{E}_{\|,inst}$. The colored segment in panel (a) is the interval chosen for the FPC analysis.}
\end{figure}

The cadence of EDP in burst mode is 8192 samples per second, while that of FPI DES for a full sky sampling is 30 ms. In order to carry out the field-particle correlation $C'_{E_{\parallel}}(v_\parallel,v_\bot,t,\tau)$ as per Equation \eqref{eq:full_cprime}, the distribution measurements and electric field must be matched in cadence. There are several ways in which this can be accomplished. We choose to average the electric field measurements that fall within the 30 ms time tags of FPI DES. This method results in a decrease in the short-timescale, peak amplitudes of the electric field as seen in \figref{fig:interval08_e_field}(a) where ``$\V{E}_{\|,inst}$, downsampled'' and ``$\V{E}_{\|,B_0}$, downsampled'' are the dowsampled values of the parallel electric field with respect to the instantaneous magnetic field and ambient magnetic field, respectively. Though the values of ``$\V{E}_{\|,B_0}$, downsampled'' used in the computation of $C'_{E_{\parallel}}(v_\parallel,v_\bot,t,\tau)$ have smaller averaged peak amplitudes, \newblue{the short-timescale peaks do provide a net force on the electrons at the precise energy-angle step timing of the FPI DES. Thus, the physical effects of the full electric field are felt by the electrons as measured in the electron velocity distribution} and are therefore included in the field-particle correlation $C'_{E_{\parallel}}(v_\parallel,v_\bot,t,\tau)$.  \figref{fig:interval08_e_field}(b) shows the difference between the downsampled $\V{E}_{\|,B_0}$ and $\V{E}_{\|,inst}$, indicating minimal $\V{E}_\bot$ contamination in $\V{E}_{\|,B_0}$. Note that the correlation $C'_{E_{\parallel}}(v_\parallel,v_\bot,t,\tau)$ in interval 08 is 62 seconds, shown as the colored segments in Figure \figref{fig:interval08_e_field}(a). This downsampling method was used for the electric field in all 20 intervals analyzed.

\acknowledgments
This work was performed under the support of University of New Hampshire contract 06-0002 under NASA Prime contract NNG04EB99C. G. G. H. was supported by NASA grants  80NSSC18K0643 and 80NSSC18K1371. A.S.A. would like to thank Jason Shuster, Narges Ahmadi, Rick Wilder, and Tak Chu Li for the useful discussions. The data used in this research is available to the public via the MMS Science Data Center (https://lasp.colorado.edu/mms/sdc/public/).\\

\bibliography{space}

\end{document}